# Employing Emotion Cues to Verify Speakers in Emotional Talking Environments

Ismail Shahin

Department of Electrical and Computer Engineering

University of Sharjah

P. O. Box  27272

Sharjah, United Arab Emirates

Tel: (971) 6 5050967

Fax: (971) 6 5050877

E-mail: ismail@sharjah.ac.ae

**Abstract**

Usually, people talk neutrally in environments where there are no abnormal talking conditions such as stress and emotion. Other emotional conditions that might affect people talking tone like happiness, anger, and sadness. Such emotions are directly affected by the patient's health status. In neutral talking environments, speakers can be easily verified; however, in emotional talking environments, speakers cannot be easily verified as in neutral talking ones. Consequently, speaker verification systems do not perform well in emotional talking environments as they do in neutral talking environments. In this work, a two-stage approach has been employed and evaluated to improve speaker verification performance in emotional talking environments. This approach employs speaker's emotion cues (text-independent and emotion-dependent speaker verification problem) based on both Hidden Markov Models (HMMs) and Suprasegmental Hidden Markov Models (SPHMMs) as classifiers. The approach is comprised of two cascaded stages that combines and integrates emotion recognizer and speaker recognizer into one recognizer. The architecture has been tested on two different and separate emotional speech databases: our collected database and Emotional Prosody Speech and Transcripts database. The results of this work show that the proposed approach gives promising results with a significant improvement over previous studies and other approaches such as emotion-independent speaker verification approach and emotion-dependent speaker verification approach based completely on HMMs.

# 1. Introduction

Listeners can obtain different types of information from speech signals. Such types of information are: 1) Speech recognition which conveys information about the content of the speech signal. 2) Speaker recognition which yields information about the speaker identity. 3) Emotion recognition that gives information about the emotional state of the speaker. 4) Health recognition which provides information on the patient's health status. 5) Language recognition that produces information of the language being spoken. 6) Accent recognition which generates information about the speaker accent. 7) Age recognition which delivers information about the speaker age. 8) Gender recognition that gives information about the speaker gender.

There are two types of speaker recognition: speaker identification and speaker verification (authentication). Speaker identification is the task of automatically determining who is speaking from a set of known speakers. Speaker verification is the task of automatically determining if a person really is the person he or she claims to be. Speaker verification can be used in intelligent health care systems [1], [2], [3], [4]. Speaker verification systems are used in hospitals which include computerized emotion categorization and assessment techniques [1]. These systems can also be used in the pathological voice assessment (functional dysphonic voices) [2]. Dysphonia is the medical term for disorders of the voice: an impairment in the ability to produce voice sounds using the vocal organs. Thus, dysphonia is a phonation disorder. The dysphonic voice can be hoarse or excessively breathy, harsh, or rough [5]. Furthermore, speaker verification systems can be used in the diagnosis of Parkinson's disease [3]. Max Little and his team at Massachusetts Institute of Technology (MIT) did some work on analyzing and evaluating the voice characteristics of patients who had been diagnosed with Parkinson's disease. They discovered that they could create a tool to detect such a disease in the speech patterns of individuals [3]. In addition, speaker verification systems can be exploited to provide assistance to multidisciplinary evaluation teams as they evaluate each child who is referred for an assessment to determine if he/she is a child with a disability and in need of special education services. The verification of children with disabilities is one of the most important aspects of both federal law and state special education regulation [4].

Speaker recognition has been an interesting research field in the last few decades, which still yields a number of challenging problems. One of the most challenging problems that face speaker recognition systems is the low performance of such systems in emotional talking environments [6], [7], [8], [9]. Emotion-based speaker recognition is one of the vital research fields in the human-computer interaction or affective computing area [10]. The foremost goal of intelligent human-machine interaction is to enable computers with the affective computing capability so that computers can verify the identity of the user in intelligent healthcare services.

## 2. Prior Work

There are many studies [11], [12], [13] that focus on speaker verification in neutral talking environments. The authors of [11] addressed the issues related to language and speaker recognition, focusing on prosodic features extracted from speech signals. Their proposed approach was evaluated using the National Institute of Standards and Technology (NIST) language recognition evaluation 2003 and the extended data task of NIST speaker recognition evaluation 2003 for language and speaker recognition, respectively. The authors of [12] described the major elements of MIT Lincoln Laboratory's Gaussian Mixture Model (GMM)-based speaker verification system in neutral talking environments. The authors of [13] focused their work on text-dependent speaker verification systems in such talking environmens. In their proposed approach, they used suprasegmental and source features, besides spectral features to verify speakers. The combination of suprasegmental, source, and spectral features significantly enhances the performance of speaker verification systems [13].

On the other hand, there is a limited number of studies [6], [7], [8], [9] that address the issue of speaker verification in emotional talking environments. The authors of [6] presented investigations into the effectiveness of the state-of-the-art speaker verification techniques: Gaussian Mixture Model-Universal Background Model and Gaussian Mixture Model-Support Vector Machine (GMM-UBM and GMM-SVM) in mismatched noise conditions. The authors of [7] examined whether speaker verification algorithms that are trained in emotional environments yield better performance when applied to speech samples obtained under stressful or emotional conditions than those trained in neutral environments only. They concluded that training of speaker verification algorithms on a broader range of speech samples, including stressful and emotional talking conditions, rather than the neutral talking condition, is a promising method to enhance speaker authentication performance [7]. The author of [8] proposed, implemented, and tested a two-stage approach for speaker verification systems in emotional talking environments based entirely on Hidden Markov Models (HMMs). He tested the proposed approach using his collected speech database and obtained 84.1% as a speaker verification performance. The authors of [9] studied the influence of emotion on the performance of a Gaussian Mixture Model-Universal Background Model (GMM-UBM) based speaker verification system in such talking environments. In their work, they proposed an emotion-dependent score normalization technique for speaker verification on emotional speech. They achieved an average speaker verification performance of 88.5% [9].

The main contribution of this work is focused on employing and evaluating a two-stage approach to verify the claimed speaker in emotional talking environments. This approach consists of two recognizers which are combined and integrated into one recognizer using both HMMs and Suprasegmental Hidden Markov Models (SPHMMs) as classifiers. The two recognizers are: emotion identification recognizer followed by speaker verification recognizer. Our present work focuses on enhancing the performance of text-independent and emotion-dependent speaker verification systems. This work deals with inter-session variability caused by different emotional states of the claimed speaker. Based on the current approach, the claimed speaker should be registered in advance in the test set (closed set). Our present work is different from one of our prior works [14] that focused on identifying speakers based on a two-stage approach. In [14], the first stage is to identify the unknown emotion and the second stage is to identify the unknown speaker given that the emotion of the unknown speaker was identified.

The motivation of this work is that speaker verification systems do not perform well in emotional talking environments as they do in neutral talking environments [6], [8], [9]. The proposed architecture of this work aims at enhancing the degraded speaker verification performance in emotional talking environments based on employing emotion cues. The present work is a continuation to the work of one of our previous studies [8] which was devoted to proposing, implementing, and testing a two-stage approach to verify speakers in emotional talking environments based completely on HMMs as a classifier and using only collected database. In addition, five extensive experiments have been conducted in the current work to assess the two-stage approach.

The remainder of this paper is organized as follows: The fundamentals of SPHMMs are covered in Section 3. Section 4 describes the two speech databases used in this work and the extraction of features. Section 5 discusses the two-stage approach and the experiments. Decision threshold is presented in Section 6. Section 7 demonstrates the results obtained in the present work and their discussion. Finally, concluding remarks are presented in Section 8.

### 3. Fundamentals of SPHMMs

SPHMMs have been developed, implemented, and evaluated by the author of [15], [16], [17] in the fields of: speaker recognition [16], [17] and emotion recognition [15]. SPHMMs have proven to be superior models over HMMs for speaker recognition in each of shouted [16] and emotional [17] talking environments.

Suprasegmental is a vocal result which expands over many sound segments in an utterance such as pitch and stress. It is usually used for tone, vowel length, and features such as nasalization. SPHMMs have the ability to summarize several states of HMMs into what is termed a suprasegmental state. Suprasegmental state can look at the observation sequence through a larger window. Such a state allows observations at rates suitable for the situation of modeling. For example, prosodic information can not be detected at a rate that is used for acoustic modeling. The prosodic features of a unit of speech are named suprasegmental features since they influence all segments of the unit. As a result, prosodic events at the levels of: phone, syllable, word, and utterance are represented by means of suprasegmental states, while acoustic events are modeled using conventional hidden Markov states.

Within HMMs, prosodic and acoustic information can be combined as given in the following formula [18],

$$log\ \mathrm{P}\left(\lambda^v, \Psi^v | \mathrm{O}\right) = (1-\alpha).\ log\ \mathrm{P}\left(\lambda^v | \mathrm{O}\right) + \alpha.\ log\ \mathrm{P}\left(\Psi^v | \mathrm{O}\right) \qquad (1)$$

where $\alpha$ is a weighting factor. When:

$$\begin{cases} 0.5 > \alpha > 0 & \text{biased towards acoustic model} \\ 1 > \alpha > 0.5 & \text{biased towards prosodic model} \\ \alpha = 0 & \text{biased completely towards acoustic model and no effect of prosodic model} \\ \alpha = 0.5 & \text{not biased towards any model} \\ \alpha = 1 & \text{biased completely towards prosodic model and no impact of acoustic model} \end{cases} \qquad (2)$$

$\lambda^v$: is the acoustic model of the $v^{th}$ emotion.

$\Psi^v$: is the suprasegmental model of the $v^{th}$ emotion.

$O$: is the observation vector or sequence of an utterance.

$P(\lambda^v | O)$ and $P(\Psi^v | O)$ can be calculated using Bayes theorem as given in Eqs. (3) and (4), respectively [19],

$$P(\lambda^v | O) = \frac{P(O | \lambda^v) P_0(\lambda^v)}{P(O)} \qquad (3)$$

$$P(\Psi^v | O) = \frac{P(O | \Psi^v) P_0(\Psi^v)}{P(O)} \qquad (4)$$

where $P_0(\lambda^v)$ and $P_0(\Psi^v)$ are the priori distribution of acoustic model and suprasegmental model, respectively.

## 4. Speech Databases and Extraction of Features

### 4.1 Collected Database

The collected speech data corpus is composed of twenty male and twenty female untrained healthy adult native speakers of American English. Untrained speakers were selected to utter sentences naturally and to avoid exaggerated expressions. Each speaker was asked to utter eight sentences where each sentence was portrayed nine times under each of the neutral, angry, sad, happy, disgust, and fear emotions. The eight sentences were unbiased towards any emotion. These sentences are:

1) *He works five days a week.*
2) *The sun is shining.*

3) *The weather is fair.*
4) *The students study hard.*
5) *Assistant professors are looking for promotion.*
6) *University of Sharjah.*
7) *Electrical and Computer Engineering Department.*
8) *He has two sons and two daughters.*

The first four sentences of this database were used in the training phase, while the last four sentences were used in the evaluation phase (text-independent experiment). The collected speech data corpus was captured in a clean environment by a speech acquisition board using a 16-bit linear coding A/D converter and sampled at a sampling rate of 16 kHz. This database is a wideband 16-bit per sample linear data. The signal samples were pre-emphasized and then segmented into frames of 16 ms each with 9 ms overlap between consecutive frames.

**4.2 Emotional Prosody Speech and Transcripts (EPST) Database**

Emotional Prosody Speech and Transcripts (EPST) data corpus was produced by Linguistic Data Consortium (LDC) [20]. This data corpus is comprised of eight professional speakers (three actors and five actresses) uttering a series of semantically neutral utterances composed of dates and numbers spoken in fifteen different emotions including the neutral state. Only six emotions were used in this work. The six emotions are: neutral, hot anger, sadness, happiness, disgust, and panic. In this database, four utterances were used in the training phase and different four utterances were used in the test phase.

## 4.3 Extraction of Features

In this work, the features that characterize the phonetic content of speech signals in the two databases are called Mel-Frequency Cepstral Coefficients (MFCCs). These coefficients have been broadly used in many studies in the fields of speech recognition [21], [22], speaker recognition [9], [23], and emotion recognition [24], [25], [26], [27], [28]. This is because such coefficients outperform other coefficients in the three fields and because they offer a high-level estimation of human auditory perception.

Most of the works [27], [29], [30] performed in the last few decades in the fields of speech recognition, speaker recognition, and emotion recognition on HMMs have been done using Left-to-Right Hidden Markov Models (LTRHMMs) because phonemes follow strictly left-to-right sequence. In this work, Left-to-Right Suprasegmental Hidden Markov Models (LTRSPHHMs) have been derived from LTRHMMs. Fig. 1 shows an example of a basic structure of LTRSPHMMs that has been obtained from LTRHMMs. In this figure, $q_1, q_2, ..., q_6$ are conventional hidden Markov states. $p_1$ is a suprasegmental state that consists of $q_1, q_2$, and $q_3$. $p_2$ is a suprasegmental state that is made up of $q_4, q_5$, and $q_6$. $p_3$ is a suprasegmental state that is composed of $p_1$ and $p_2$. $a_{ij}$ is the transition probability between the $i^{th}$ conventional hidden Markov state and the $j^{th}$ conventional hidden Markov state. $b_{ij}$ is the transition probability between the $i^{th}$ suprasegmental state and the $j^{th}$ suprasegmental state.

In this work, the number of conventional states of LTRHMMs, $N$, is six. The number of mixture components, $M$, is ten per state, with a continuous mixture observation density is selected for these models. In LTRSPHMMs, the number of suprasegmental states is two. Therefore, each three conventional states of LTRHMMs are summarized into one suprasegmental state.

## 5. Speaker Verification Based on the Two-Stage Approach and the Experiments

Given a registered speaker talking in *m* emotions, the overall proposed approach to verify the claimed speaker based on his/her emotion cues is shown in Fig. 2. The aim of the two-stage approach is to deal with inter-session variability caused by different emotional states of the claimed speaker. Fig. 2 shows that the overall two-stage architecture is comprised of two cascaded stages. The two stages are:

**Stage *a*: Emotion Identification**

The first stage of the overall approach is to identify the unknown emotion that belongs to the claimed speaker (emotion identification problem). In this stage, *m* probabilities are computed based on SPHMMs and the maximum probability is chosen as the identified emotion as given in the following formula,

$$E^* = \arg\max_{m \geq e \geq 1} \left\{ P\left(O \mid \lambda^e, \Psi^e\right) \right\} \quad (5)$$

where,

$E^*$: is the index of the identified emotion.

$O$: is the observation sequence of the unknown emotion that belongs to the claimed speaker.

$P(O | \lambda^e, \Psi^e)$: is the probability of the observation sequence $O$ of the unknown emotion that belongs to the claimed speaker given the $e^{th}$ SPHMM emotion model ($\lambda^e$, $\Psi^e$).

The $e^{th}$ SPHMM emotion model has been derived in the training phase for every emotion using the forty speakers generating all the first four sentences with a repetition of nine utterances per sentence. Therefore, the total number of utterances used to derive each SPHMM emotion model in this phase is 1440 (40 speakers × 4 sentences × 9 utterances / sentence). SPHMM training phase is very similar to conventional HMM training phase. In SPHMM training phase, suprasegmental models are trained on top of acoustic models of HMMs. A block diagram of this stage is illustrated in Fig. 3.

**Stage *b*: Speaker Verification**

The next stage of the two-stage approach is to verify the speaker identity based on HMMs given that his/her emotion was identified in the previous stage (emotion-specific speaker verification problem) as given in the following formula,

$$\Lambda(O) = \log\left[P\left(O | E^*\right)\right] - \log\left[P\left(O | \overline{E}^*\right)\right] \qquad (6)$$

where,

*Λ(O)*: is the log-likelihood ratio in the *log* domain.

$P(O|E^*)$: is the probability of the observation sequence $O$ that belongs to the claimed speaker given the true identified emotion.

$P(O|\bar{E}^*)$: is the probability of the observation sequence $O$ that belongs to the claimed speaker given the false identified emotion. Eq. (6) shows that the likelihood ratio is computed between model trained using data from claimed speaker and recognized emotion.

The probability of the observation sequence $O$ that belongs to the claimed speaker given the true identified emotion can be computed as [31],

$$log\ P(O|E^*) = \frac{1}{T}\sum_{t=1}^{T} log\ P(o_t|E^*) \tag{7}$$

where, $O = o_1 o_2 \ldots o_t \ldots o_T$.

The probability of the observation sequence $O$ that belongs to the claimed speaker given the false identified emotion can be computed using a set of $B$ imposter emotion models: $\{\bar{E}_1^*, \bar{E}_2^*, \ldots, \bar{E}_B^*\}$ as,

$$log\ P(O|\bar{E}^*) = \left\{\frac{1}{B}\sum_{b=1}^{B} log\ [P(O|\bar{E}_b^*)]\right\} \tag{8}$$

where $P(O|\bar{E}_b^*)$ can be computed using Eq. (7). The value of $B$ in this work is equal to $6 - 1 = 5$ emotions. Fig. 4 demonstrates a block diagram of this stage.

In the evaluation phase, each one of the forty speakers used nine utterances per sentence of the last four sentences (text-independent) under each emotion. The total number of utterances used in this phase is 8640 (40 speakers × 4 sentences × 9 utterances / sentence × 6 emotions). In this work, 34 speakers (17 speakers per gender) are used as claimants and the rest of speakers are used as imposters.

# 6. Decision Threshold

Two types of error can take place in a speaker verification problem, namely, false rejection (miss probability) and false acceptance (false alarm probability). When a valid identity claim is rejected, it is called a false rejection error; on the other hand, when the identity claim from an imposter is accepted, it is named a false acceptance.

Speaker verification problem based on emotion identification requires making a binary decision based on two hypotheses. Hypothesis $H_0$ if the claimed speaker belongs to a true emotion or hypothesis $H_1$ if the claimed speaker comes from a false emotion.

The log-likelihood ratio in the log domain can be defined as,

$$\Lambda(O) = log\left[P\left(O|\lambda_C, \Psi_C\right)\right] - log\left[P\left(O|\lambda_{\overline{C}}, \Psi_{\overline{C}}\right)\right] \qquad (9)$$

where,

$O$: is the observation sequence of the claimed speaker.

$\lambda_C, \Psi_C$ : is the SPHMM claimant emotion model.

$P(O|\lambda_C, \Psi_C)$: is the probability that the claimed speaker belongs to a true identified emotion.

$\lambda_{\overline{C}}, \Psi_{\overline{C}}$ : is the SPHMM imposter emotion model.

$P(O|\lambda_{\overline{C}}, \Psi_{\overline{C}})$: is the probability that the claimed speaker comes from a false identified emotion.

The last step in the verification process is to compare the log-likelihood ratio with the threshold (θ) in order to accept or reject the claimed speaker, *i.e.,*

$$\text{Accept the claimed speaker if } \Lambda(O) \geq \theta$$

$$\text{Reject the claimed speaker if } \Lambda(O) < \theta$$

Open set speaker verification often uses thresholding to make a decision if a speaker is out of the set. Both types of error in speaker verification problem rely on the threshold used in the decision making process. A tight value of threshold makes it difficult for false speakers to be falsely accepted but at the expenditure of falsely rejecting true speakers. On the other hand, a loose value of threshold facilitates true speakers to be accepted continually at the expense of falsely accepting false speakers. In order to set a proper value of threshold that meets with a desired level of a true speaker rejection and a false speaker acceptance, it is essential to know the distribution of true speaker and false speaker scores. An acceptable process for setting a value of threshold is to assign a loose initial value of threshold and then let it adjust by setting it to the average of up-to-date trial scores. This loose value of threshold gives inadequate protection against false speaker trials.

## 7. Results and Discussion

In the current work, a two-stage approach based on both HMMs and SPHMMs as classifiers has been employed and tested using separately the collected and EPST databases when $\alpha = 0.5$ for speaker verification in emotional talking environments. This specific value of $\alpha$ has been chosen to avoid biasing towards either acoustic or prosodic model.

Table 1 and Table 2 show confusion matrices of stage $a$ using the collected and EPST databases, respectively. The two matrices represent percentage of confusion of the unknown emotion with the other emotions based on SPHMMs. Table 1 (for example) demonstrates the following:

1. The most easily recognizable emotion is neutral (99%). Hence, the performance of verifying speakers talking neutrally is the highest compared to that of verifying speakers talking in other emotions as shown in Table 3 (least percentage Equal Error Rate, EER) using the same database.

2. The least easily recognizable emotion is angry (86%). Therefore, speaker verification performance when speakers talk in angry emotion is the least compared to that when speakers talk in other emotions as shown in Table 3 (highest percentage EER) using the same database.

3. Column 3 (angry emotion), for example, shows that 2% of the utterances that were portrayed in angry emotion were evaluated as produced in neutral state, 3% of the utterances that were uttered in angry emotion were recognized as generated in sad emotion. This column shows that angry emotion has the highest confusion percentage with disgust emotion (6%). Therefore, angry emotion is highly confusable with disgust emotion. The column also shows that angry emotion has the least confusion percentage with happy emotion (1%).

Table 3 shows percentage Equal Error Rate (EER) in emotional talking environments based on the two-stage framework using each of the collected and EPST databases when $\alpha = 0.5$. This table indicates that the average value of EER using the collected database is 7.75%, while the average value of EER using EPST database is 8.17%. The table shows that the least value of EER happens when the claimed speaker speaks neutrally, while the highest value of EER occurs when the claimed speaker talks in angry emotion. This table shows that the percentage EER under all emotions, except under the neutral state, is high. This high percentage EER may be attributed to the following reasons:

1. The identified emotion of the claimed speaker has not been perfectly identified. The average emotion identification performances based on SPHMMs are 91.67% and 92.15% using the collected and EPST databases, respectively.

2. The verification stage (stage *b*) produces another system degradation performance in addition to the degradation in emotion identification performance. This is because some claimants are rejected as imposters and some imposters are accepted as claimants. Therefore, the given EER in Table 3 is the resultant of the EER of both stage *a* and stage *b*. Since the performance of emotion identification stage is imperfect, the two-stage framework could have a negative impact on the overall performance especially when the emotion in stage *a* has been falsely identified.

The authors of [9] achieved an average EER of 11.48% in emotional talking environments using GMM-UBM based on emotion-independent method. In the present work, the achieved average EER based on the two-stage approach is less than that obtained based on their method [9]. The author of [8] obtained 15.9% as an average EER in emotional talking environments based on HMMs only. It is evident that the attained results of average EER based on the two-stage approach are less than those achieved in [8].

Five extensive experiments have been conducted in this work to assess the achieved results based on the two-stage architecture. The five experiments are:

(1) Percentage EER based on the two-stage approach is compared with that based on an emotion-independent speaker verification approach using separately the collected and EPST databases. The obtained average EER using the emotion-independent approach based on HMMs only and using each of the collected and EPST databases is given in Table 4. Based on this table, the average value of EER using the collected and EPST databases is 14.75% and 14.58%, respectively.

A statistical significance test has been performed to show whether EER differences (EER based on the two-stage framework and that based on the emotion-independent approach) are real or simply due to statistical fluctuations. The statistical significance test has been carried out based on the Student's *t* Distribution test as given in the following formula,

$$t_{1,2} = \frac{\bar{x}_1 - \bar{x}_2}{SD_{pooled}} \qquad (10)$$

where,

$\bar{x}_1$: is the mean of the first sample of size *n*.

$\bar{x}_2$: is the mean of the second sample of the same size.

SD $_{pooled}$: is the pooled standard deviation of the two samples given as,

$$SD_{pooled} = \sqrt{\frac{SD_1^2 + SD_2^2}{2}} \qquad (11)$$

where,

SD$_1$: is the standard deviation of the first sample of size *n*.

SD$_2$: is the standard deviation of the second sample of the same size.

In this work, $\bar{x}_{3,collect} = 7.75, SD_{3,collect} = 2.91, \bar{x}_{3,EPST} = 8.17,$ $SD_{3,EPST} = 3.14, \bar{x}_{4,collect} = 14.75, SD_{4,collect} = 4.28, \bar{x}_{4,EPST} = 14.58,$ $SD_{4,EPST} = 4.14$. These values have been calculated using Table 3 (collected and EPST databases) and Table 4 (collected and EPST databases), respectively. Based on these values, the calculated *t* value using the collected database of both Tables 3 and 4 is $t_{4,3 \text{ (collected)}} = 1.913$ and the calculated *t* value using EPST database of both Tables 3 and 4 is $t_{4,3 \text{ (EPST)}} = 1.745$. Each calculated *t* value is higher than the tabulated critical value at *0.05* significant level $t_{0.05} = 1.645$. Therefore, the conclusion that can be drawn in this experiment states that the two-stage speaker verification approach outperforms the emotion-independent speaker verification approach. Therefore, inserting emotion identification stage into speaker verification system in emotional talking environments significantly enhances speaker verification performance compared to that without such a stage.

(2) In stage *a*, the *m* probabilities are computed based on SPHMMs. To compare the impact of using acoustic features on emotion identification (stage *a*) with that using suprasegmental features, Eq. (5) has become as,

$$E^* = \arg\max_{m \geq e \geq 1} \left\{ P\left(O \mid \lambda^e\right) \right\} \qquad (12)$$

Therefore, the *m* probabilities in this experiment are computed based on HMMs. The obtained percentage EER employing emotion cues based on the two-stage approach and using HMMs only in both stage *a* and stage *b* using the collected and EPST databases is given in Table 5. The average value of EER using the collected and EPST databases is 15.58% and 14.50%, respectively.

In this experiment, $\bar{x}_{5,\text{collect}} = 15.58$, $SD_{5,\text{collect}} = 3.85$, $\bar{x}_{5,\text{EPST}} = 14.50$, $SD_{5,\text{EPST}} = 3.71$. Based on these values, the calculated *t* value of both Tables 3 and 5 using the collected database is $t_{5,3 \text{ (collect)}} = 2.294$ and the calculated *t* value of the two tables using EPST database is $t_{5,3 \text{ (EPST)}} = 1.842$. Each calculated *t* value is larger than $t_{0.05} = 1.645$. Therefore, the percentage EER based on using SPHMMs in stage *a* is lower than that based on using HMMs in the same stage. It can be concluded from this experiment that SPHMMs are superior to HMMs for speaker verification in emotional talking environments.

Fig. 5 and Fig. 6 show Detection Error Trade-offs (DETs) curves using the collected and EPST databases, respectively. Each curve compares speaker verification in emotional talking environments based on the two-stage approach with that based on the emotion-independent approach. These two figures evidently demonstrate that the two-stage approach is superior to the emotion-independent approach for speaker verification in emotional talking environments.

(3) The two-stage approach has been evaluated for different values of $\alpha$. Fig. 7 and Fig. 8 show average percentage EER based on the two-stage framework for different values of $\alpha$ (0.0, 0.1, …, 0.9, 1.0) using the collected and EPST databases, respectively. The two figures indicate that increasing the value of the weighting factor has a significant effect on minimizing EER and hence improving speaker verification performance in emotional talking environments (excluding the neutral state) based on the two-stage architecture. Therefore, it is apparent, based on this architecture, that suprasegmental hidden Markov models have more influence on speaker verification performance in such talking environments than acoustic hidden Markov models. These two figures also show that the least percentage EER takes place when the classifiers are entirely biased towards suprasegmental models and no impact of acoustic models ($\alpha = 1$).

(4) The two-stage approach has been assessed for the worst case scenario. This scenario takes place when stage *b* receives false input (false identified emotion) from stage *a*. The average percentage EER for the worst case scenario based on SPHMMs when $\alpha = 0.5$ is 15.11% and 15.25% using the collected and EPST databases, respectively. These values are very close to those attained using the one-stage approach (14.75% and 14.58% using the collected and EPST databases, respectively). It can be concluded from this experiment that the percentage EER for the worst case scenario based on the two-stage approach is very close to that based on the one-stage approach.

(5) An informal subjective assessment of the two-stage approach has been performed with ten nonprofessional listeners (human judges) using the collected speech data corpus. A total of 960 utterances (20 speakers × 2 genders × 6 emotions × the last 4 sentences of the database) have been used in this assessment. During the evaluation, each listener is asked two separate questions for every test utterance. The two questions are: identify the unknown emotion and verify the claimed speaker provided the unknown emotion was identified. The average emotion identification performance and the average speaker verification performance is 90.5% and 88.14%, respectively.

## 8. Concluding Remarks

This work employed and evaluated a two-stage approach that combines and integrates emotion recognizer and speaker recognizer into one recognizer using both HMMs and SPHMMs as classifiers to enhance speaker verification performance in emotional talking environments. Several experiments have been separately carried out in such environments using two different and separate speech databases. Some conclusions can be drawn from this work. Firstly, the emotional state of the speaker has a negative impact on speaker verification performance. Secondly, the significant improvement of speaker verification performance in emotional talking environments based on the two-stage approach reveals promising results of such an approach. Thirdly, emotion-dependent speaker verification architecture is superior to emotion-independent speaker verification architecture (one-stage approach). Therefore, emotion cues significantly contribute in alleviating the deteriorated speaker verification performance in these talking environments. Fourthly, suprasegmental hidden Markov models outperform conventional hidden Markov models for speaker verification systems in such talking environments. Furthermore, the highest speaker verification performance happens when the classifiers are completely biased towards suprasegmental models and no influence of acoustic models. Finally, the two-stage recognizer performs almost the same as the one-stage recognizer when the second stage (stage *b*) receives false identified emotion from the first stage (stage *a*).

There are some limitations in this work. First, the processing computations and the time consumed in the two-stage approach are slightly greater than those in the one-stage approach. Second, the two-stage approach requires all emotions of the claimed speaker to be available to the system in the training phase. Hence, the two-stage architecture is restricted to a closed set case. Finally, speaker verification performance based on the two-stage approach is limited. This is because the performance of the overall approach is a resultant of the performances of both stage *a* and stage *b*. Since the performance of each stage is imperfect, the overall performance is consequently imperfect.

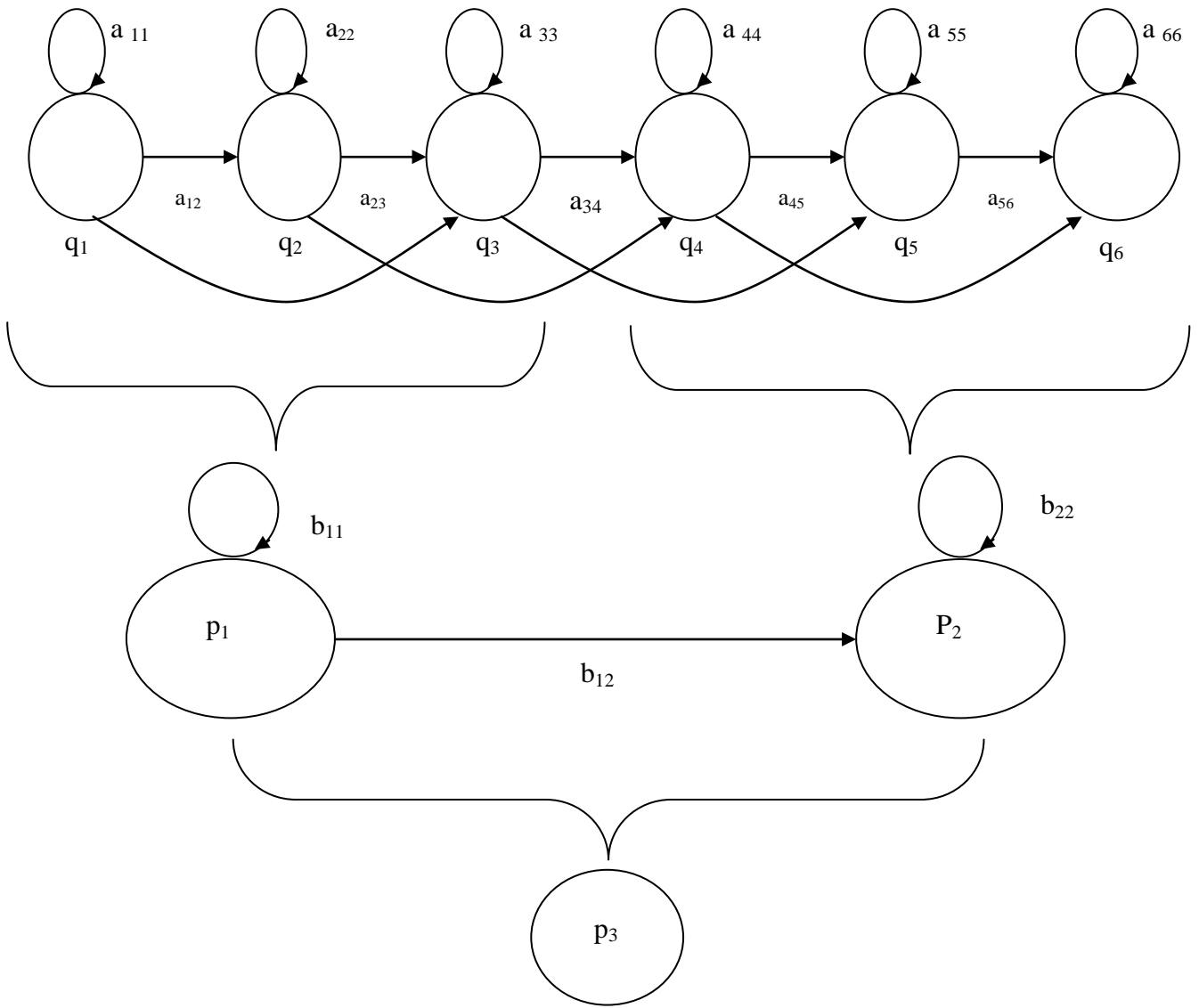

**Fig. 1.** Basic structure of LTRSPHMMs

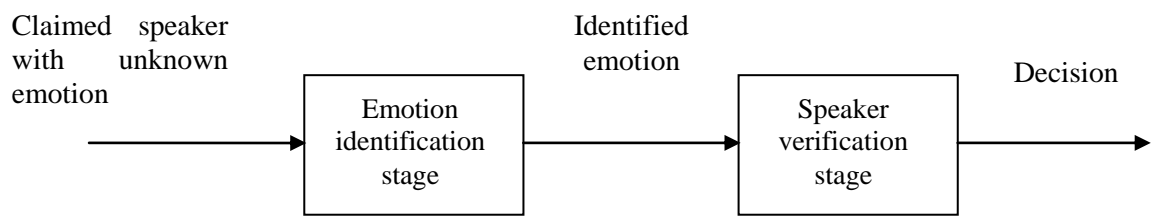

**Fig. 2.** Block diagram of the overall two-stage approach

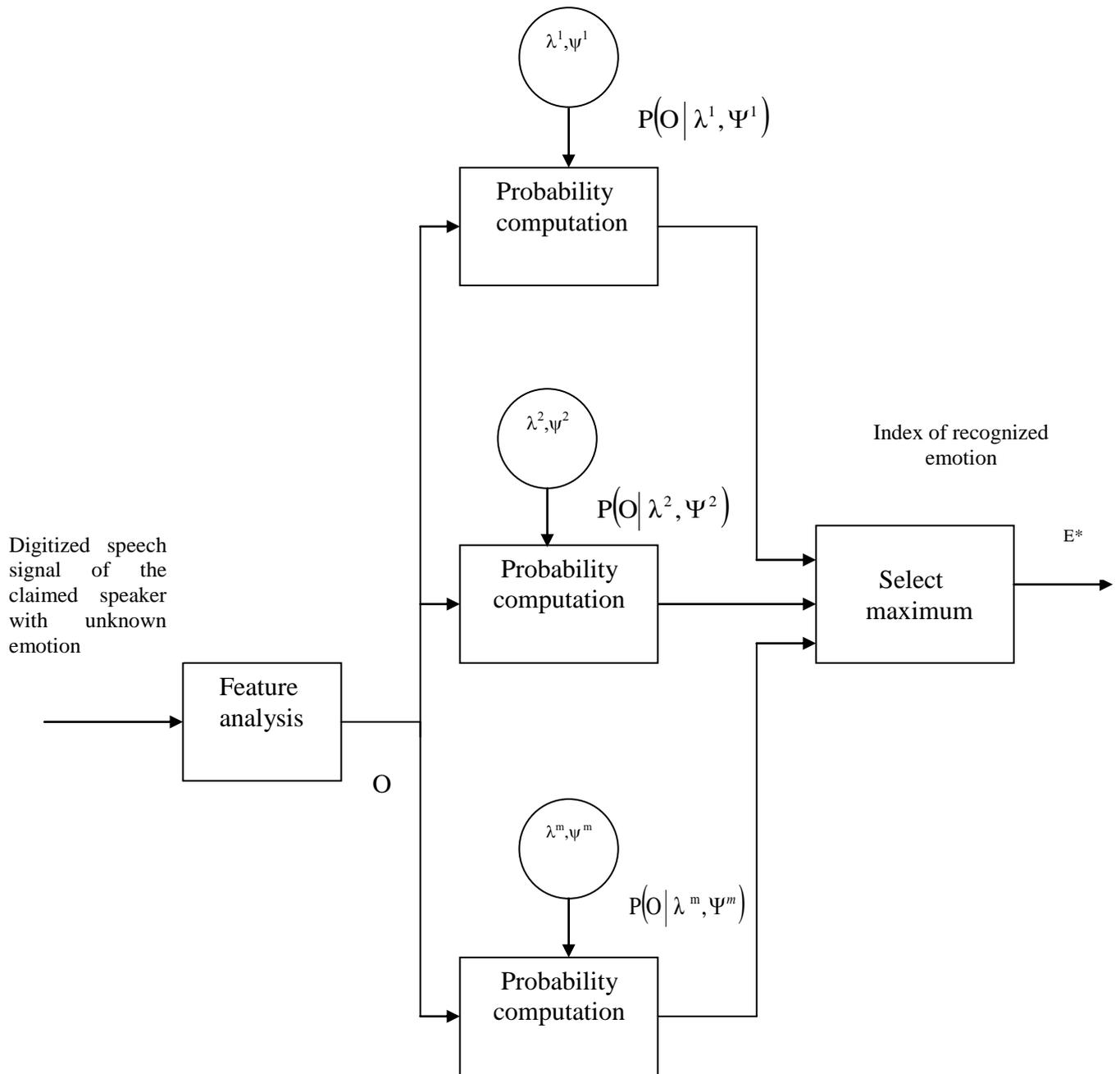

**Fig. 3.** Block diagram of stage *a* of the overall two-stage approach

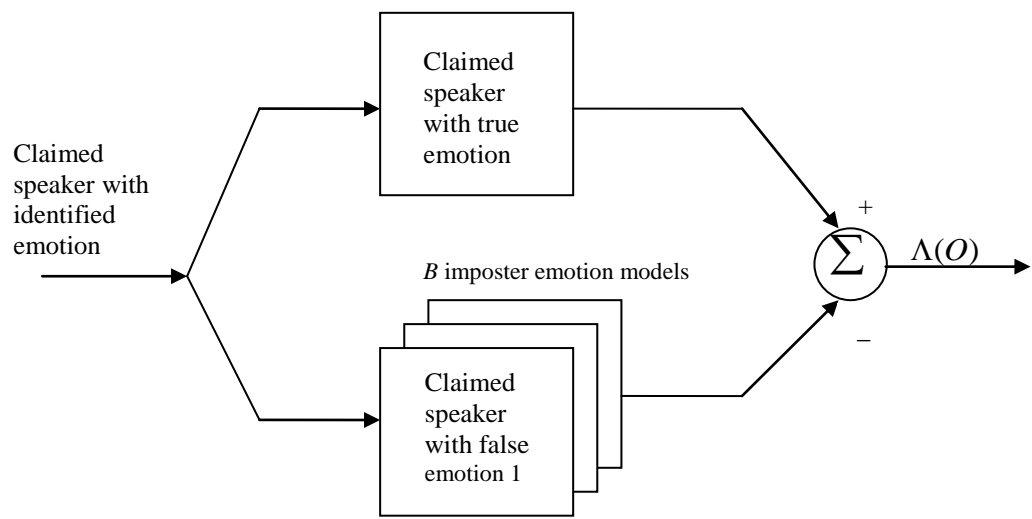

**Fig. 4.** Block diagram of stage *b* of the overall two-stage approach

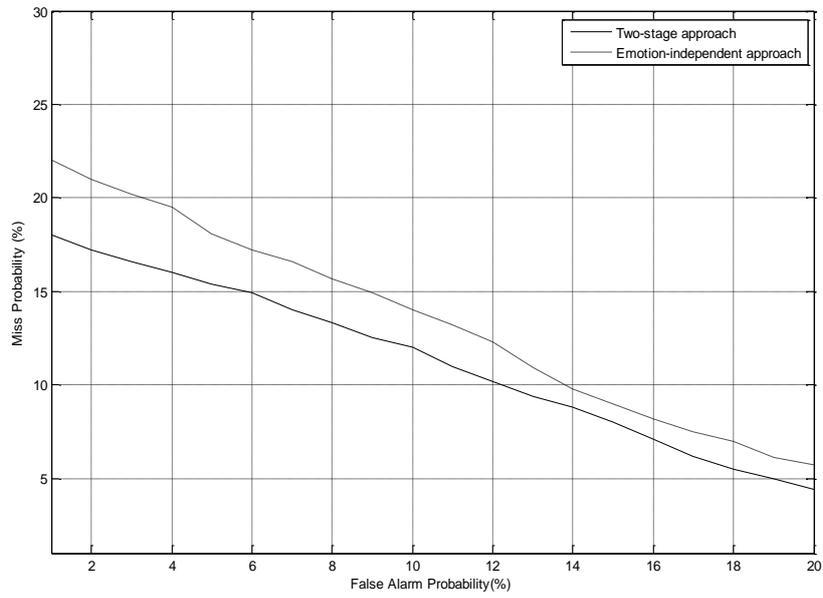

**Fig. 5.** DET curve based on each of the two-stage and emotion-independent approaches using the collected database

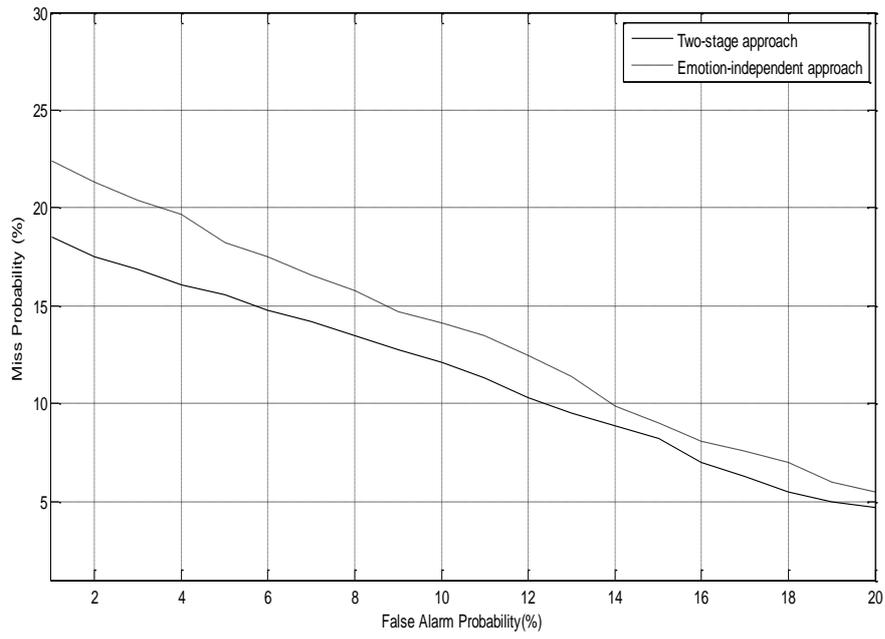

**Fig. 6.** DET curve based on each of the two-stage and emotion-independent approaches using EPST database

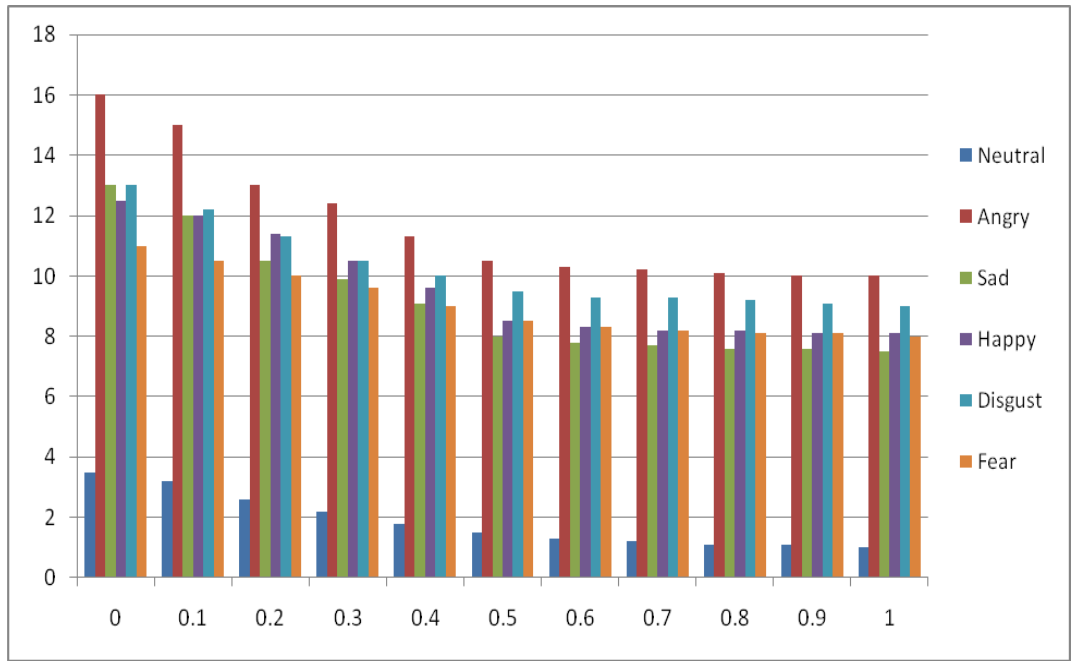

**Fig. 7.** Average EER (%) versus ($\alpha$) based on the two-stage approach using collected database

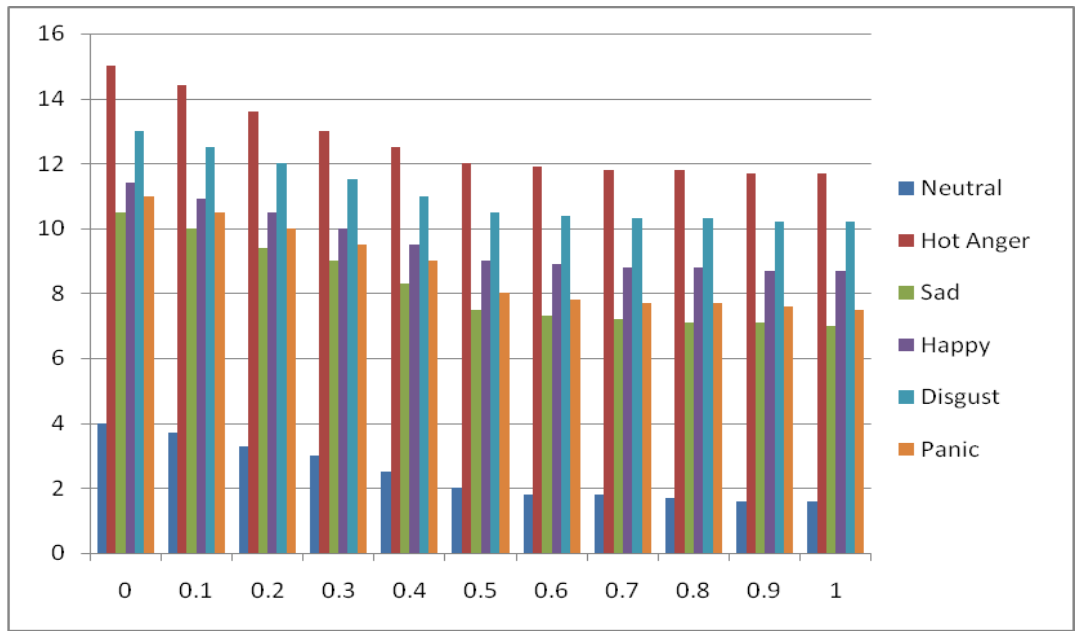

**Fig. 8.** Average EER (%) versus ($\alpha$) based on the two-stage approach using EPST database

Table 1

Confusion matrix based on SPHMMs of stage *a* using collected database when

$\alpha = 0.5$

|  | Percentage of confusion of the unknown emotion with other emotions | | | | | |
|---|---|---|---|---|---|---|
| Model | Neutral | Angry | Sad | Happy | Disgust | Fear |
| Neutral | **99%** | 2% | 1% | 1% | 0% | 0% |
| Angry | 0% | **86%** | 1% | 1% | 3% | 5% |
| Sad | 0% | 3% | **96%** | 2% | 4% | 1% |
| Happy | 0% | 1% | 0% | **92%** | 1% | 1% |
| Disgust | 0% | 6% | 0% | 2% | **87%** | 3% |
| Fear | 1% | 2% | 2% | 2% | 5% | **90%** |

Table 2

Confusion matrix based on SPHMMs of stage *a* using EPST database when

$\alpha = 0.5$

| | Percentage of confusion of the unknown emotion with other emotions | | | | | |
|---|---|---|---|---|---|---|
| Model | Neutral | Hot Anger | Sad | Happy | Disgust | Panic |
| Neutral | **99%** | 4% | 1% | 3% | 1% | 1% |
| Hot Anger | 0% | **88%** | 1% | 1% | 4% | 4% |
| Sad | 0% | 1% | **97%** | 1% | 3% | 1% |
| Happy | 1% | 1% | 0% | **94%** | 0% | 0% |
| Disgust | 0% | 5% | 0% | 1% | **86%** | 3% |
| Panic | 0% | 1% | 1% | 0% | 6% | **91%** |

Table 3

Percentage equal error rate based on the two-stage approach using the collected

and EPST databases when $\alpha = 0.5$

|  | EER (%) | |
| --- | --- | --- |
| Emotion | Collected database | EPST database |
| Neutral | 1.5 | 2 |
| Angry/Hot Anger | 10.5 | 12 |
| Sad | 8 | 7.5 |
| Happy | 8.5 | 9 |
| Disgust | 9.5 | 10.5 |
| Fear/Panic | 8.5 | 8 |

Table 4

Percentage equal error rate based on the emotion-independent approach using the collected and EPST databases

|  | EER (%) | |
| --- | --- | --- |
| Emotion | Collected database | EPST database |
| Neutral | 6 | 6 |
| Angry/Hot Anger | 18.5 | 18 |
| Sad | 13.5 | 13.5 |
| Happy | 15.5 | 15.5 |
| Disgust | 16.5 | 16.5 |
| Fear/Panic | 18.5 | 18 |

Table 5

Percentage equal error rate based on HMMs only in both stage *a* and stage *b* using

the collected and EPST databases

| Emotion | EER (%) | |
|---|---|---|
| | Collected database | EPST database |
| Neutral | 8 | 7 |
| Angry/Hot Anger | 20.5 | 18.5 |
| Sad | 15.5 | 14.5 |
| Happy | 15 | 14 |
| Disgust | 16.5 | 15.5 |
| Fear/Panic | 18 | 17.5 |